# Atomic-scale Structural and Chemical Characterization of Hexagonal Boron Nitride Layers Synthesized at the Wafer-Scale with Monolayer Thickness Control

Wei-Hsiang Lin[1], Victor W. Brar [1,2,3], Deep Jariwala[1,4], Michelle C. Sherrott[1,4], Wei-Shiuan Tseng[5], Chih-I Wu [6], Nai-Chang Yeh[2, 5], and Harry A. Atwater[1, 2,4*]

* haa@caltech.edu

1. Thomas J. Watson Laboratory of Applied Physics, California Institute of Technology, Pasadena, CA 91125, United States

2. Kavli Nanoscience Institute, California Institute of Technology, Pasadena, CA 91125, United States

3. Department of Physics, University of Wisconsin-Madison, Madison, WI 53711, United States

4. Resnick Sustainability Institute, California Institute of Technology, Pasadena, CA 91125, USA.

5. Department of Physics, California Institute of Technology, Pasadena, CA 91125, United States

6. Graduate Institute of Photonics and Optoelectronics and Department of Electrical Engineering, National Taiwan University, Taipei, Taiwan, Republic of China

**Abstract**

Hexagonal boron nitride (h-BN) is a promising two-dimensional insulator with a large band gap and low density of charged impurities that is isostructural and isoelectronic with graphene. Here we report the chemical and atomic-scale structure of CVD-grown wafer-scale (~25 cm$^2$) h-BN sheets ranging in thickness from 1-20 monolayers. Atomic-scale images of h-BN on Au and graphene/Au substrates obtained by scanning tunneling microscopy (STM) reveal high h-BN crystalline quality in monolayer samples. Further characterization of 1-20 monolayer samples indicates uniform thickness for wafer-scale areas; this thickness control is a result of precise control of the precursor flow rate, deposition temperature and pressure. Raman and infrared spectroscopy indicate the presence of B-N bonds and reveal a linear dependence of thickness with growth time. X-ray photoelectron spectroscopy (XPS) shows the film stoichiometry, and the B/N atom ratio in our films is $1 \pm 0.6\%$ across the range of thicknesses. Electrical current transport in metal/insulator/metal (Au/h-BN/Au) heterostructures indicates that our CVD-grown h-BN films can act as excellent tunnel barriers with a high hard-breakdown field strength. Our results suggest that large-area h-BN films are structurally, chemically and electronically uniform over the wafer scale, opening the door to pervasive application as a dielectric in layered nanoelectronic and nanophotonic heterostructures.

**Introduction**

Since the isolation of graphene on an insulating substrate[1-7], a variety of other layered materials have been isolated and characterized, and have opened up an exciting field of research[8-11]. However the electronic quality of two-dimensional (2D) active layers in devices is highly sensitive to their immediate environment owing to the large surface to volume ratio. Therefore a crystalline 2D material that can serve the role of an insulating substrate, encapsulating layer or gate dielectric is highly desirable in device applications. Hexagonal boron nitride (h-BN) has emerged as a promising material for these applications. It has been observed experimentally that encapsulating other 2D materials in h-BN not only enhances the performance of devices but also extends their durability to long time-scales[12-13]. However most previous studies have used mechanically exfoliated h-BN derived from bulk crystals which does not allow for control over layer thickness and sample lateral size. While numerous reports about chemical vapor deposition (CVD) synthesis of h-BN already exist[14-32] ranging from low pressure CVD (LPCVD) for monolayer growth to ambient pressure CVD (APCVD) for multilayer growth, most of these approaches do not yield precise thickness control from the monolayer scale to thick multilayers over large areas. Many previous experiments have reported on use of solid ammonia-borane as the boron and nitrogen source. The sublimation of a solid source gives poor control of precursor flow rate and partial pressure in the growth chamber over large length scales. Further, monolayer to sub-2nm growth of h-BN is a process catalyzed by the Cu surface and requires low growth pressures. However the catalytic activity of Cu surface is diminished after growth of 1-2 nm h-BN on its surface and thus growth of thicker h-BN films occurs solely via van der Waals epitaxy, requiring higher growth pressures. No system has yet been developed that can achieve both of these growth condition regimes; therefore most prior work on CVD grown h-BN has been able to control both BN thickness and lateral thickness uniformity over technologically relevant areas. Our CVD methods features precursor flow and pressure control systems which combines the merits of both LPCVD and APCVD by allowing precise control over the precursor flow rate and partial pressure of the precursor in the growth zone over a wide range of growth chamber pressure and temperatures. This permits growth of high quality mono to sub 5-layer h-BN films on a Cu foil which require low growth pressures as well as thicker h-BN films on Cu foils, by switching to APCVD mode.

We image the atomic structure of monolayer h-BN/Au sheets and monolayer h-BN/graphene/Au heterostructures using scanning tunneling microscopy (STM). Atomic STM images of CVD-grown monolayer h-BN/Au film monolayers and h-BN/graphene heterostructures indicate high crystalline quality. The crystal structure and chemical composition of the resulting h-BN film are also characterized by atomic force microscopy (AFM), Raman spectroscopy, infrared transmission measurements and x-ray photoelectron spectroscopy (XPS). The electrical properties of the thin and thick h-BN films were systematically measured on metal-insulator-metal (MIM) tunneling devices to understand the dielectric properties of these films. Our results demonstrate a new standard and state-of-the-art for large area synthesis of h-BN potentially enabling applications in nanoelectronic and nanophotonic devices.

**Results and Discussion**

To grow ultrathin (< 1.5 nm) h-BN films, we use a growth pressure of ~ 2 Torr. Figure 1b shows an optical photograph and atomic force microscopy (AFM) analysis of a wafer-scale h-BN monolayer which has transferred onto a 285 nm $SiO_2$/Si substrate. Raman spectra acquired over six different spots on the sample (denoted by x), as seen in the adjacent plot suggests that the sample thickness and quality is uniform over the entire $cm^2$ scale. Likewise, the growth mode can be switched from a slow growth rate of ~0.3 nm/min in catalytically controlled CVD at low pressures to a high growth rate of 1 nm/min at near atmospheric pressures (~500 Torr). Figure 1c illustrates the large area uniformity of ~ 15 monolayer h-BN as inferred from the optical micrograph and corresponding Raman spectra. To validate the precise control of layer thickness and growth uniformity, a series of growth experiments were performed for varying times at both low and ambient pressures to produce h-BN films of varying layer number/thicknesses. These h-BN films were then characterized with several spectroscopic techniques to characterize their structure and chemical composition. Atomic scale structure and areal homogeneity of h-BN films were revealed by scanning tunneling microscopy (STM) for CVD-grown monolayer h-BN. Despite numerous reports on CVD synthesis of h-BN, little is known about atomic scale electronic structure for layers grown on polycrystalline Cu foils, due to the roughness of the Cu foil substrate which renders STM difficult. Also, the film thickness inhomogeneity seen in most prior reports[33-37] would prevent image formation by direct electron tunneling through h-BN. Notably, h-BN is difficult to characterize using STM due to its insulating character. To enable adequate sample conductance, many research groups have used a graphene/ h-BN heterostructure which exploits the conductance of graphene to visualize the atomic structure of the h-BN layer underneath the graphene. In contrast, we are able to use STM to directly image our CVD-grown monolayer h-BN films on Au (111)/mica substrates transferred using polymer free transfer method[38]. Figure 2a is a schematic of the STM measurement configuration for monolayer h-BN sheets on Au (111) substrates. Figure 2b shows a representative STM image of the monolayer h-BN sheet on Au (111) without any post transfer annealing. The atomically-resolved h-BN honeycomb structure is clearly visible, superimposed on the Au (111) herringbone reconstruction pattern. The appearance of the distorted hexagonal lattice in the STM image indicates strong surface tension originating from the interaction between boron nitride atoms and the Au (111) substrate herringbone reconstruction. Figure 2c shows schematic of STM measurements of the monolayer h-BN sheet on monolayer graphene/Au (111) substrate and Figure 2d, e show topographic STM images acquired from two different areas of the single-layer h-BN on graphene. Both the atomic lattices of h-BN and longer range moiré patterns can be clearly revealed. The moiré pattern is formed by interference between the h-BN layer and underlying graphene/Au (111) substrate, and can be attributed to their lattice mismatch (a = 0.252 nm, b = 0.246 nm) and rotational misalignment. The h-BN and graphene sheets interact through van der Waals forces, and display the same topographic conformal mapping to the underlying Au (111). However, the relative rotation angle between the graphene and h-BN sheets can be modified by tiny wrinkles and bubbles are inevitably introduced during the transfer process of h-BN onto graphene as seen in Fig. 2d (left) and original grain boundary of h-BN and graphene. Changing the rotation angles between the h-

BN and graphene lattices, leads to moiré patterns with different periodicities and orientations as observed in Figure 2d and e (center). The periodicity of moiré pattern presented in Figure 2d is 3.4 nm and 2.0 nm in Figure 2e. The twist angle between the h-BN and graphene lattices can be ascertained from the structure of the Moire pattern and is given by

$$\theta = cos^{-1}[1 - \frac{a^2b^2 - \lambda^2(b-a)^2}{2ab\lambda^2}]$$

where a is the h-BN lattice constant, b is the graphene lattice constant, and $\lambda$ is the periodicity of the moiré pattern. Thus, the twist angle $\theta$ of Figure 2d and 2e is found to be (4 ± 0.1) ° and (7 ± 0.1) °, respectively. Alternatively, the twist angle $\theta$ can also be extracted by performing a fast Fourier transform (FFT) analysis of the STM images, as shown in the inset of Figure 2d and 2e. The outside set of spots corresponds to the reciprocal lattice of h-BN, while the inner set of spots are assigned to the moiré pattern stemming from the rotation between the monolayer h-BN and graphene substrate. The atomic resolution STM images presented here can be achieved in image locations over a wide area of the wafer-scale sample, indicating excellent h-BN sheet surface quality for layers transferred with the polymer-free[38] transfer method and supporting the existence of high crystalline quality in CVD-grown h-BN down to the monolayer level. The STM results provide information about the atomic scale structure and crystallinity of monolayer h-BN. However they do not provide information about macroscopic-scale h-BN film thickness homogeneity, composition or structure.

Raman spectra of an h-BN layer typically have two active $E_{2g}$ modes, one at 1366 cm$^{-1}$ which is strong and corresponds to vibrations of B and N moving against each other in the plane and another at 51.8 cm$^{-1}$, which is attributed to sliding between whole planes. However the lower frequency mode is more difficult to observe because of proximity to the Rayleigh diffusion, as well as the presence of a fluorescence background. The width, intensity and position of these Raman features are sensitive to h-BN thickness, and these dependencies were determined by combining Raman measurement with STM results and AFM results. We use monolayer h-BN as a thickness calibration for every run and verify the integrated intensity of this calibration point with the value marked with an arrow in Figure 3b (Top). The integrated intensity of the h-BN layers determined using this protocol gives a reasonable estimate of the layer thickness and matches very well with the estimates given by the AFM line profiles (see supporting information S1). Figure 3a shows Raman spectra of monolayer h-BN to 15 layers h-BN films, showing the Raman intensity increases with the number of layers. The integrated intensity is plotted as a function of the number of layers in Figure 3b (Top). Figure 3b (Bottom) suggests that the peak position have a redshift as layers decrease and the FWHM become shaper as the layer number decreases. While Raman spectroscopy suggests the presence of B-N bonds and a linear dependence of thickness with growth time, they do not provide any information on the stoichiometry of the grown films. To probe the chemical composition, we used X-ray photoelectron spectroscopy (XPS) to determine the B/N ratio. Figure 3c shows the XPS spectra of as-grown h-BN films on a Cu foils with a film thickness varying from monolayer to 30 layers. It has been previously reported[39] that bulk boron nitride with hexagonal phase exhibits a B 1s core level at 190.1 eV. Figure 3c (left) shows XPS B1s core level spectra with a peak center at 190.2 eV, which is very close to the h-BN bulk phase value. Figure 3c (right) shows that the N 1s peak is located at 397.7 eV, similar to the reported position of the N1s spectrum (398.1 eV) for h-BN. Both the B 1s and N 1s spectra indicate that the configuration for B and N atoms is the B-N bond, implying that the hexagonal phase is the phase of our BN films. Further, it can also be seen that the intensities of the B1s and N1s increases with increasing layer thickness of our films. In addition, we also observe that the intensity of the Cu2p peak weakens with the increasing thickness of the h-BN films (see supporting information S3) further corroborating our precise thickness controlled growth. Finally, quantitative analysis of the B1s and N1s spectra indicates that the B/N atom ratio in our films was 1 ± 0.6% across the range of thicknesses. These results evidently confirm growth of high quality h-BN layer and continuous film on Cu foil using ammonia borane by our optimized CVD process. The large area uniformity and high crystalline quality of our CVD grown h-BN films makes them ideal for applications as ultrathin dielectrics in optoelectronic devices[38, 40-41]. Therefore, to evaluate the dielectric strength and leakage through our h-BN films, we investigate the electronic properties of tunnel junctions in which h-BN acts as a barrier layer between two gold electrodes. The dielectric properties of the CVD h-BN films with different thicknesses were measured by fabricating metal/h-BN/metal (MIM) capacitors and measuring current–voltage (I-V) characteristics. Figure 4a shows a schematic diagram of an Au/h-BN/Au capacitor. We use template stripped gold films as our bottom electrode with root mean square (RMS) roughness less than 0.5nm. The h-BN films of various thicknesses were transferred onto the template stripped gold. Then, we used standard electron beam deposition techniques to deposit 100nm gold through a shadow mask to define the top electrodes. The contact area was 10μm×10μm. Figure 4b shows I-V measurements of h-BN layers with various thicknesses from 1 to 15 layers. Mono-, bi-, and 4-layer samples show measurable low-bias conductance, which we ascribe to direct tunneling. Thicker samples are insulating at low bias and show sharp increases at a breakdown voltage that increases with thickness. The inset in the Figure 4b shows the conductance as a function of sample thickness, which decays exponentially, as expected for direct tunneling. The current densities at the two metal electrodes and through the h-BN layers of different thicknesses were investigated as a function of voltage, as plotted in Figure 4c. As shown in Figure 4c, the measured currents of the thin h-BN films agreed well with the Poole-Frenkel (PF) emission model, indicating that a trap-assisted PF emission mechanism dominated the transport mechanism for the leakage current in our h-BN films. The Figure 4c shows the PF plot using the following equation:

$$I(V)_{PF} = AqN_c\mu V d \exp\left[\frac{-q(\Phi_T - \sqrt{\frac{qVd}{\pi\varepsilon_0\varepsilon_r}})}{kT}\right]$$

where A, q, $N_c$, μ, $\Phi_T$, V, d and h are the effective area, electron charge, density of state in the conduction band, electronic mobility in the oxide, trap energy level in the h-BN, voltage, h-BN thickness and Planck's constant respectively. Finally, for thicker (> 1 nm) h-BN films we performed irreversible dielectric breakdown measurements to determine the hard-breakdown voltage (see supporting information S 4) and corresponding field strength of the ultrathin h-BN. Figure 4 (d) plots the breakdown field strength as a function of the h-BN thickness. In h-BN films with a thicknesses less than 5 nm, the breakdown voltage increased linearly with h-BN thickness, indicating very high quality films at the few-layer limit. Breakdown field strength approaching ~ 4.3 MV/cm were observed for 4.5 nm thick BN films.

**Conclusions**

We have imaged the atomic-scale structure of monolayer h-BN sheets on Au, and moiré patterns on monolayer h-BN/graphene heterostructures using scanning tunneling microscopy (STM). Atomic STM images of monolayer h-BN film and moiré patterns on monolayer h-BN/graphene heterostructures show the high crystalline quality of the CVD grown h-BN up to the atomic level. We also introduced a hybrid LP and APCVD system that uses controlled precursor to grow uniform, layer by layer thickness controlled wafer scale h-BN films with thicknesses ranging from monolayer to 10 nm. Spectroscopic characterization suggests that the films are stoichiometric and highly uniform over wafer-scale areas. Electrical measurements for metal-insulator-metal (Au/h-BN/Au) structures indicate that our CVD-grown h-BN films can act as an excellent tunnel barrier with a high hard-breakdown field strength. Successful large area CVD growth of h-BN films defines a new state-of-the-art for application of this material in future large-area, electronic and photonic devices.

**Methods**

**Pre-treatment of copper foil.** Copper foil (25 μm, 99.999% pure, Alfa Aesar, item no. 10950) was soaked and sonicated in acetone and isopropyl alcohol (IPA) for 30 min consecutively to remove organic impurities. Then, it was washed with deionized water and dried with nitrogen gas.

**Synthesis of mono- and multilayer h-BN.** The h-BN films were grown using a home-built hybrid CVD setup. Figure 1a shows the schematic of the growth setup. The setup comprises of a 52 mm inner diameter (I.D.) horizontal split tube furnace (MTI Corporation). The solid precursor ammonia-borane ($NH_3$-$BH_3$) powder, (97% purity, Sigma-Aldrich) is contained in a home-made quartz container, attached to the main growth chamber (22 mm I.D. quartz tube) via a leak valve and heated separately from the quartz tube via use of a resistive heating belt. Cu foils (25 μm, 99.999% pure, Alfa Aesar) are used as the catalytic growth substrates. The pressure in the growth chamber can be independently controlled via an angle valve at the vacuum pump while the pressure in the precursor bubbler can be controlled via the leak valve and carrier gas flow rates. The monolayer and multilayer h-BN was synthesized using a pressure controllable CVD system. The copper foil was inserted into the center of a 22 mm I.D. quartz tube, heated by a horizontal split-tube furnace. The Ammonia borane ($NH_3$-$BH_3$) (97% purity, from Sigma-Aldrich), stable in an atmospheric environment, was used as the precursor. It was loaded in a homemade quartz container which is isolated from the main CVD system with a leak valve to control the flow rate. The quartz tube inlet and outlet were blocked by the filters to prevent the BN nanoparticles from diffusing into the gas line. First, the quartz tube was pump down to $5\times10^{-3}$ torr, and then ultrahigh purity grade hydrogen gas was introduced during the temperature ramp-up of the furnace (pressure ~100mtorr, flow rate ~50 sccm). The copper foil was annealed at 950°C in hydrogen for 60 min to obtain a smooth surface. After annealing, the ultrahigh purity argon gas (300 sccm) was introduced into the system and waited for 30min to stable the tube environment. The precursor was heated to 130°C and decomposed to hydrogen gas, monomeric aminoborane, and borazine gas. After the precursor temperature reach 130°C, the manual valve between quartz tube outlet and the pump was slowly closed and stopped until the pressure reach 20 torr. When the desired pressure was achieved, the leak valve to the precursor was open. The typical growth time is 3 min for monolayer h-BN layer and 20 min for the 20 nm thickness h-BN layer. To atomically control the thickness of h-BN layer, it is very important to have a leak valve to control the flow rate of the precursor. Also, we can change the growth rate of the h-BN by changing the pressure of the growth environment. After growth, the tube furnace was cooled down with the cooling rate ~16°C/min.

**Transfer of mono- and multilayer h-BN.** To transfer h-BN onto a target substrate, the conventional poly(methyl methacrylate) (PMMA) transfer method and polymer-free transfer method [41] were applied based on different purposes.

**Characterization.** Atomic Force Microscopy (AFM) (Bruker Dimension Icon), were done using tapping mode to characterize the surface morphology of the h-BN film transferred on the SiO$_2$/Si substrate. The quality of the h-BN film was characterized using the Raman spectroscopy, X-ray photoemission spectroscopy (XPS) and Scanning tunneling microscopy. The Raman spectra were taken with a Renishaw M1000 micro-Raman spectrometer system using a 514.3 nm laser (2.41 eV) as the excitation source. A 50X objective lens with a numerical aperture of 0.75 and a 2400 lines/mm grating were chosen during the measurement to achieve better signal-to-noise ratio.

XPS was performed under 10$^{-9}$ torr with a Surface Science Instruments M-Probe instrument utilizing Al K$_\alpha$ X-rays and a hemispherical energy analyzer. STM was carried out with an Omicron system at room temperature.

**Acknowledgements**

The authors gratefully acknowledge support from the Department of Energy, Office of Science under Grant DE-FG02-07ER46405 (W.H.L. and H.A.A.) and for use of facilities of the DOE "Light-Material Interactions in Energy Conversion" Energy Frontier Research Center (DE-SC0001293). D.J. and M.C.S. acknowledge additional support from Resnick Sustainability Institute Graduate and Postdoctoral Fellowships whereas V.W.B. acknowledge additional support from the Kavli

Nanoscience Postdoctoral Fellowship. The authors thank Prof. George Rossman for access to the Raman and FTIR tools. The authors thank I-Te Lu for the discussion about the moiré pattern calculation. The authors acknowledge support from the Beckman Institute of the California Institute of Technology to the Molecular Materials Research Center.

**Competing financial interests**

The authors declare no competing financial interests.

**Figure 1.**

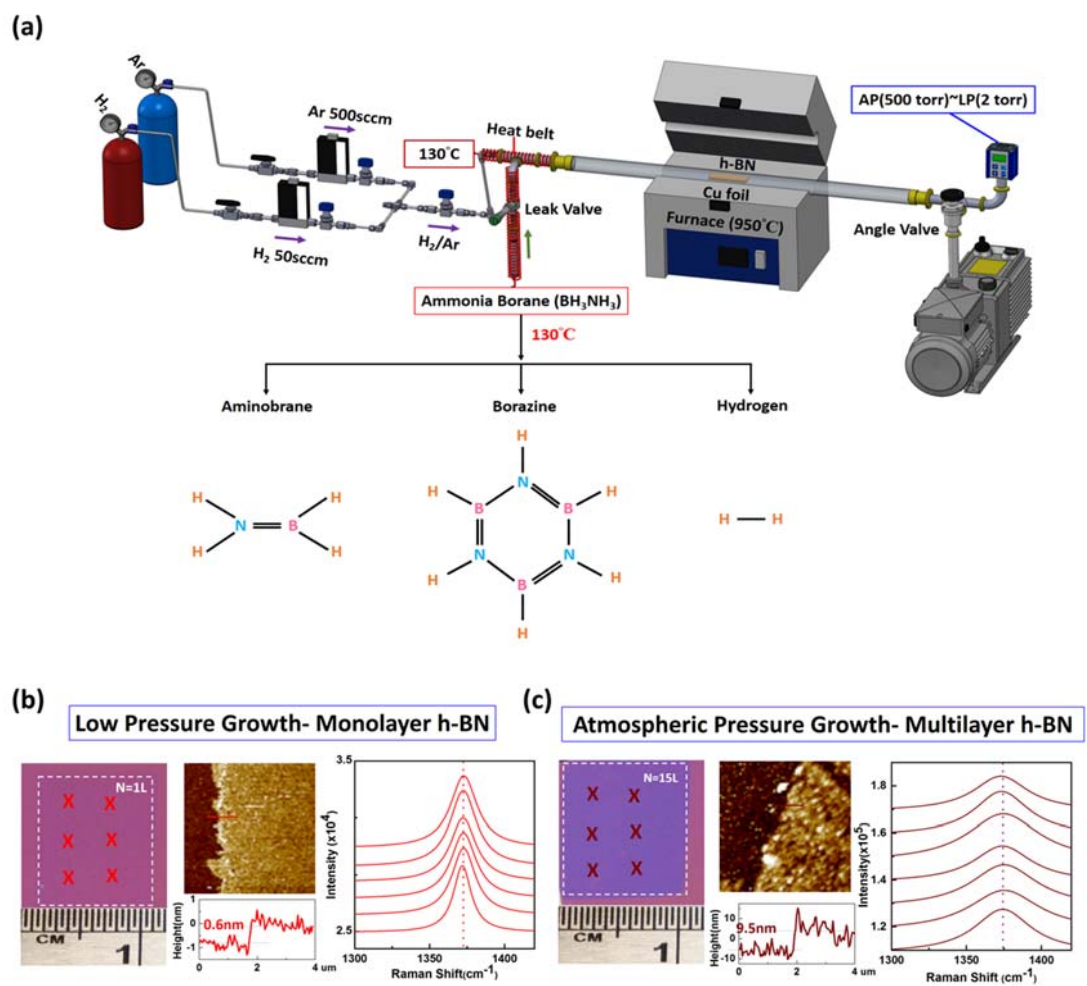

Figure 1. (a) Schematic diagram of hybrid atmospheric pressure and low pressure CVD system used for h-BN grow. (b) Photograph of a large and uniform monolayer h-BN film on a 285 nm thick SiO$_2$/Si substrate, and corresponding AFM image and Raman spectra. (c) Photograph of a large and uniform multilayer h-BN film on a 285 nm thick SiO$_2$/Si substrate, and corresponding AFM image and Raman spectra.

**Figure 2.**

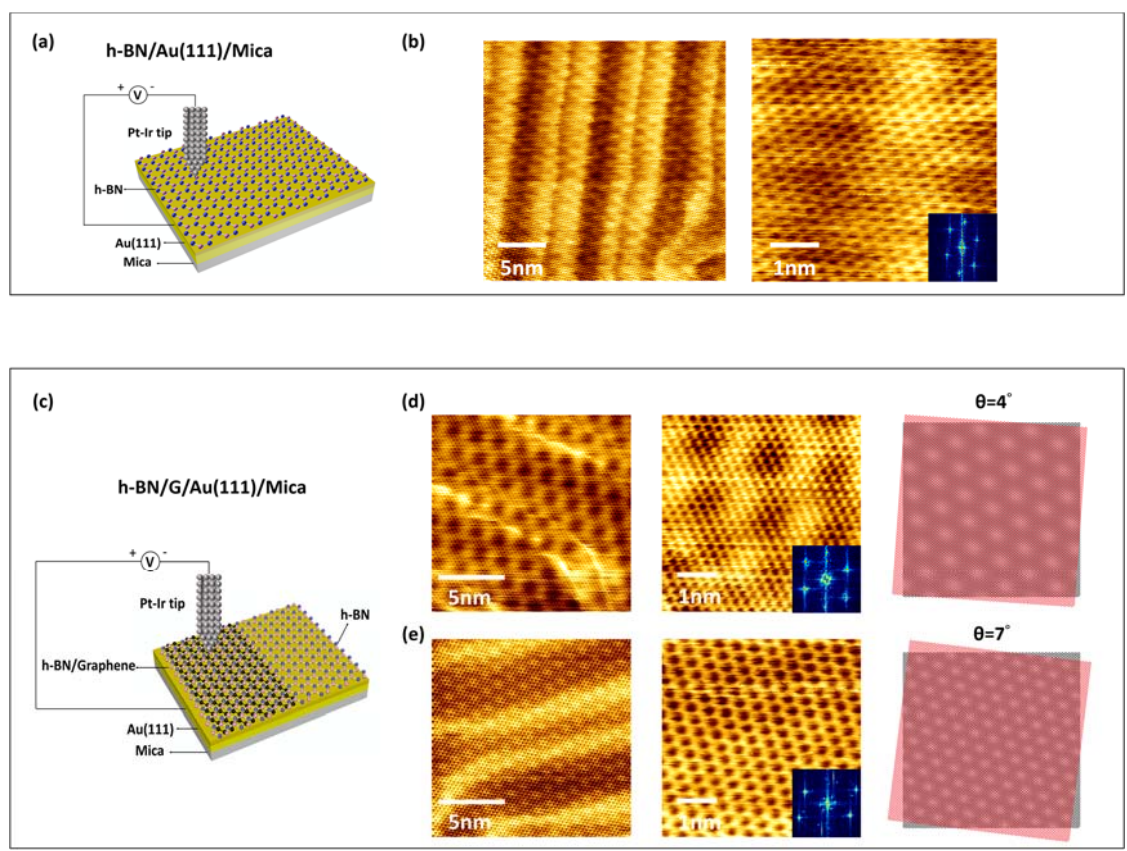

Figure 2. (a) STM measurement schematics on monolayer h-BN film. (b) The representative STM image of the h-BN after transfer to the Au (1 1 1) substrate, with $V_{sample}$=0.5 V and $I_{tunnel}$= 0.5 nA. The image size is 25 nm×25 nm (Left) and 5 nm × 5 nm (Right). (c) STM measurement schematics on monolayer h-BN/graphene heterostructure. (d) Topography of a moiré superlattice with periodicity of 3.7nm and a 4 ° twist between h-BN and graphene. The image was acquired under $V_{sample}$=0.5 V and $I_{tunnel}$= 0.5 nA and image size is 15 nm ×15 nm (Left) and 5 nm × 5 nm (Right). (e) Topography of a moiré superlattice with periodicity of 2.0 nm and a 7 ° twist between h-BN and graphene. The image was acquired under $V_{sample}$= 0.5 V and $I_{tunnel}$= 0.5 nA and image size is 15 nm ×15 nm (Left) and 5 nm × 5 nm (Right). The insets of (d) and (e) are Fourier transform patterns of corresponding STM images.

**Figure 3.**

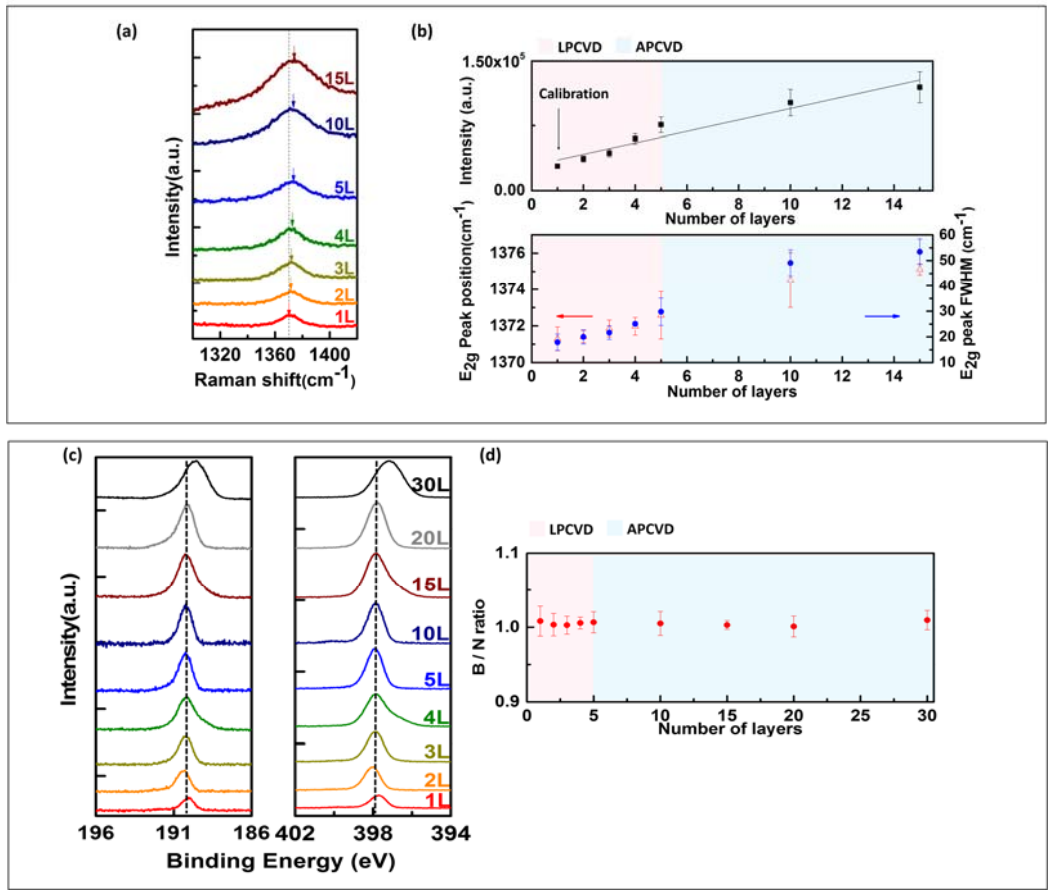

Figure 3. (a) Raman spectra for h-BN layers with 1-15 atomic layers. (b) (Top) Integrated intensity shows a steady increase with increase in the layer number of h-BN. (b) (Bottom left) The position of $E_{2g}$ peak vs. number of monolayers of h-BN, showing the blue shift with increased number of h-BN layers. (Bottom right) The Full width at half maximum vs. number of h-BN layers, showing a steady increase of the FWHM with h-BN thickness. (c) X-ray photoelectron spectra (XPS) from different thickness of h-BN layers on Cu foil. (d) High resolution B1s and N1s peaks corresponding to the thickness from 1 ~ 30 layers h-BN films on Cu foil. (Right) Quantitative analysis of the B1s and N1s spectra indicates that the B/N atom ratio in our films was 1 ± 0.6% across the range of thicknesses.

**Figure 4.**

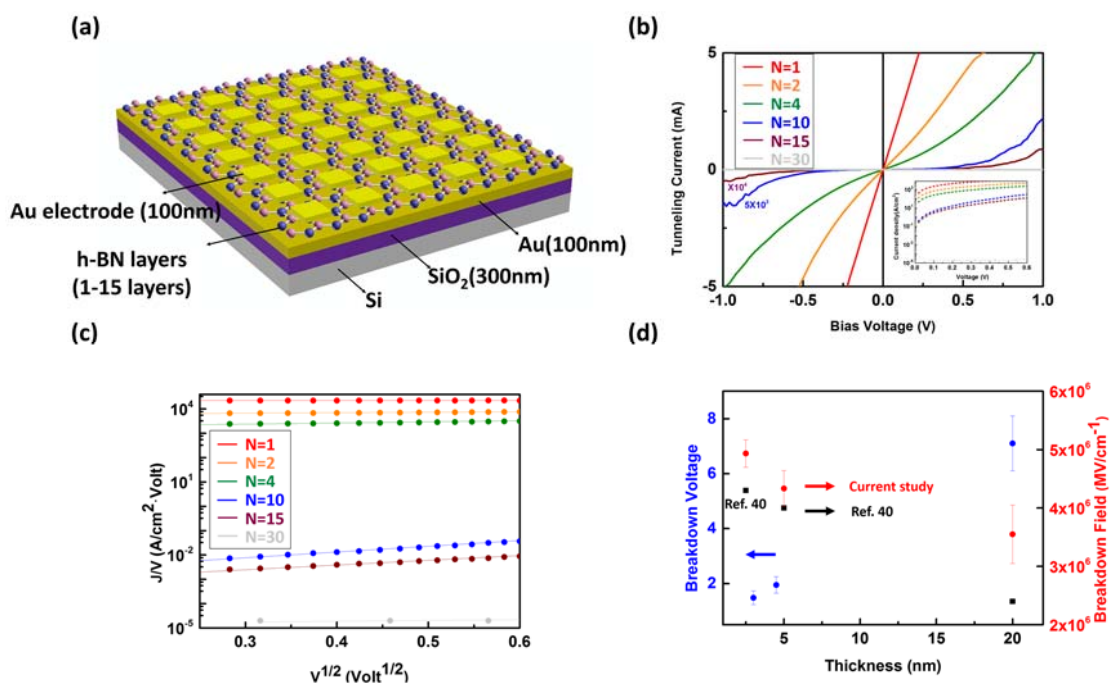

Figure 4. (a) A schematic diagram of the Au / h-BN / Au (MIM) capacitors fabricated on a Si substrate. (b) Characteristic I−V curves for Au / h-BN / Au devices with different thicknesses of BN insulating layer: red curve, monolayer of BN; orange, bilayer; green, four layer; navy, 10 layers; and purple, 15 layers. The inset of (a) is typical J-V characteristics of a MIM capacitor, described by the field-assisted tunneling model. The h-BN thickness range was less than 5 nm. The inset shows a PF emission plot (J/V versus $1/V^{1/2}$). (d) The breakdown characteristics as a function of the h-BN film thickness.

**Supporting information**

AFM images and photographs of bi-layer h-BN to 30-layer h-BN on SiO$_2$/Si substrate and the height profile, infrared phonon of h-BN with the film thickness ranging from 30 layers to 1 layer, X-ray photoelectron spectra - Cu2p spectrum of 1 ~ 30 layers h-BN films, irreversible dielectric breakdown measurements to determine the hard-breakdown voltage, film thickness with respect to the Borazine partial pressure, growth mechanism of the h-BN synthesis by APCVD and LPCVD, SEM images showing triangular ad-layer with different domain size, cross – section TEM image of 5-layer h-BN, and AFM image of 5-layer h-BN with 0.69 nm root mean square roughness.

**TOC.**

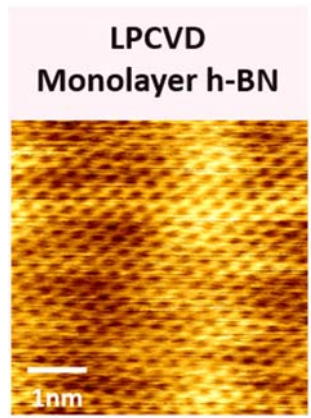 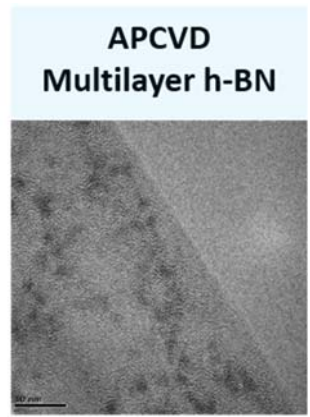